\documentclass[12pt]{article}
\usepackage{epsfig}
\usepackage{latexsym}
\begin{document}
\title{Gauge fields beyond perturbation theory.}
\author{ A.A.Slavnov \thanks{E-mail:$~~$ slavnov@mi.ras.ru}
\\Steklov Mathematical Institute, Russian Academy
of Sciences\\ Gubkina st.8, GSP-1,119991, Moscow\\ and Moscow State
University} \maketitle

\begin{abstract}
A new formulation of nonabelian gauge theories, introducing new
ghost fields and new symmetry is proposed. This formulation does not
suffer from Gribov ambiguity and allows to quantize nonabelian gauge
fields beyond perturbation theory.
\end{abstract}

\section{Introduction}
 I am going to discuss a new approach to nonabelian gauge theories,
which allows to quantize them unambigously, provides a gauge
invariant infrared regularization of these theories and may lead to
a possibility of finding solitons in the topologicaly nontrivial
sectors of nonabelian gauge theories.

Quantum Chromodynamics (QCD) is considered as theoretical basis of
strong interaction physics. No experimental facts contradicting QCD
were discovered. At the same time from the point of view of the
theory QCD is far from being completed. A consistent theory of color
confinement is absent. Even the quantization of nonabelian gauge
fields beyond perturbation theory strictly speaking does not exist .

\section{A general scheme of the method.}

Progress in physics was usually related to the introduction of new
symmetries.

Recent examples are given by gauge theories. QED may be formulated
in terms of the stress tensor, depending only on the electric and
magnetic fields, however much more transparent formulation is
presented by the quantization in a manifestly covariant gauge. But
using a covariant gauge we inevitably introduce unphysical
excitations, corresponding to temporal and longitudinal photons.
Simultaneously the theory acquires a new symmetry, gauge invariance.
This invariance provides decoupling of unphysical excitations and
guarantees unitarity of the scattering matrix in the space of
transversal photons. Yang-Mills theory became really popular only
after its formulation in the Lorentz covariant terms and explicit
proof of its renormalizability. The gauge invariance of the Higgs
model allows to give a manifestly renormalizable theory describing a
massive gauge theory.

In this talk I wish to make a propaganda for a class of symmetries,
which were introduced in my paper rather long ago \cite{Sl1} , but
recently were applied successfully to the nonperturbative
quantization of non-Abelian gauge theories, construction of the
infrared regularization, applicable beyond perturbation theory.

This symmetry is based on the equivalence theorems. It is well known
that the physical content of the theory does not change under
canonical transformations. The same statement with some reservations
related to the renormalization properties is true also for point
transformations $\varphi=\varphi'+f(\varphi')$

One can also consider more general transformations, which contain
explicitely the time derivatives of the fields. Let us transform the
fields as follows
\begin{equation}
\varphi= \frac{\partial^n \varphi'}{\partial t^n}+
f(\frac{\partial^{n-1}\varphi'}{\partial t^{n-1}}, \ldots
\frac{\partial \varphi'}{\partial t})= \tilde{f}(\varphi')
 \label{1a}
\end{equation}
The spectrum is obviously changed under this transformation. New
unphysical excitations appear. The question about the unitarity of
the transformed theory arises.

Some ideas about possible violations of unitarity by this
transformation are given by the path integral representation for the
scattering matrix
\begin{equation}
S= \int \exp \{i\int L(\varphi)dx\}d\mu(\varphi); \quad
lim_{t\rightarrow \pm \infty}\varphi(x)= \varphi_{out,in}(x)
 \label{1b}
\end{equation}
If the change (\ref{1a}) does not change the asymptotic conditions,
then the only effect of such transformation is the appearance of a
nontrivial jacobian
\begin{equation}
L(\varphi)\rightarrow \tilde{L}(\varphi')=L[\varphi(\varphi')]+
\bar{c}^a \frac{\delta \varphi^a}{\delta \varphi'^b} c^b
 \label{1c}
\end{equation}
For all new excitations one should take the vacuum boundary
conditions.But it is by no means obvious that such boundary
conditions may be imposed. To answer this question we note that the
transformed lagrangian (\ref{1c}) is invariant with respect to a new
symmetry
\begin{eqnarray}
\delta \varphi'_a=c_a \varepsilon\nonumber\\
\delta c_a=0; \quad \delta \bar{c}_a= \frac{\delta L}{\delta
\varphi_a}(\varphi')\varepsilon
 \label{1d}
\end{eqnarray}
In these equations $\varepsilon$ is a constant anticommuting
parameter. On mass shell these transformations are nilpotent and
generate a conserved charge $Q$, belonging to the Grassmann algebra.
In this case there exists an invariant subspace of states
annihilated by $Q$, which has a semidefinite norm. (\cite{Sl1}). For
asymptotic space this condition reduces to
\begin{equation}
Q_0|\phi>_{as}=0 \label{1e}
\end{equation}
The scattering matrix is unitary in the subspace which contains only
excitations of the original theory. However the theories described
by the $L$ and the $\tilde{L}$ are different, and only expectation
values of the gauge invariant operators coincide.

 A very nontrivial
generalization is obtained if one transforms the $\tilde{L}$ further
shifting the fields $\varphi$ in the topologicaly trivial sector by
constants. It is not an allowed change of variables in the path
integral as it changes the asymptotic of the fields. The unitarity
of the "shifted" theory is not guaranteed and a special proof (if
possible) is needed.

Using this method one can construct a renormalizable formulation of
nonabelian gauge theories free of ambiguity.

In fact it is not necessary to introduce higher derivatives.
Necessary ingredients are new ghost excitations, and new symmetry of
the Lagrangian.

These ideas were successfully implemented in the papers (\cite{Sl2},
\cite{QS1}, \cite{QS2}, \cite{Sl3})
A problem of unambiguos quantization of nonabelian gauge theories
beyond perturbation theory originates from the classical theory:
Even in classical theory the equation
\begin{equation}
D_{\mu}F_{\mu \nu}=0 \label{a}
\end{equation}
does not determine the Cauchi problem. To deal with gauge theory one
has to impose the gauge condition, selecting a unique representative
in a gauge equivalent class.

Differential gauge conditions: $L(A_{\mu}, \varphi)=0 \rightarrow$
which contains a differential operator as we shall see lead to
appearance of Gribov ambiguity. One can try to avoid this problem by
applying so called algebraic gauge conditions:
$\tilde{L}(A_{\mu},\varphi)=0$. The most known condition of this
kind is so called Hamiltonian gauge $A_0=0$. However these gauges
also lead to problems. From practical point of view the most
important problem is the absence of a manifest Lorentz invariance.

Let us consider the problem of ambiguity for the case of Cou\-lomb
gauge. To answer the question about ambiguity in the choice of a
representative in the class of gauge invariant configurations in the
case of the Cou\-lomb gauge, we must consider a possibility of
existence of several solutions of the equation $\partial_iA_i=0$.
\begin{eqnarray}
\partial_iA_i=0\nonumber\\
A_i'=(A^\Omega)_i\nonumber\\
\triangle \alpha^a+ig \varepsilon^{abc}\partial_i(A_i^b \alpha^c)=0
\label{1f}
\end{eqnarray}
The last equation has nontrivial solutions rapidly decreasing at
spatial infinity, therefore the Cou\-lomb gauge does not select a
unique representative among gauge equivalent configurations. This
fact was firstly noticed by V.N.Gribov \cite{Gr} and later
generalized by I.Singer \cite{Si} to arbitrary gauge. I wish to
emphasize that in perturbation theory the only solution of the
(eq.\ref{1f})is $\alpha=0$. So in perturbation theory the problem of
ambiguity is absent. There are two possibilities to solve the
problem of ambiguity :

1. Use of this phenomenon to try to explain confinement e.t.c.
(Series of works by D.Zwanzi\-ger \cite{Zw} and others.)

2.To avoid the Gribov problem by using new (equivalent) formulation
of the Yang-Mills theory using more ghost fields. In the following I
consider in more details the second option.

\section{Formulation of the Yang-Mills theory free of Gribov ambiguity}

Let us consider the classical ($SU(2)$)Lagrangian
\begin{eqnarray}
L=- \frac{1}{4}F_{\mu \nu}^aF_{\mu \nu}^a  -m^{-2}(D^2
\tilde{\phi})^*
(D^2\tilde{\phi}) +(D_{\mu}e)^* (D_{\mu}b)+(D_{\mu}b)^*(D_{\mu}e)\nonumber\\
+ \alpha^2(D_{\mu} \tilde{\phi})^*( D_{\mu}
\tilde{\phi})-\alpha^2m^2(b^*e+e^*b) \label{2a}
\end{eqnarray}
where $\phi$ is a two component complex doublet, and
\begin{equation}
\tilde{\phi}= \phi- \hat{\mu}; \quad \hat{\mu}=(0, \mu
\sqrt{2}g^{-1}) \label{2}
\end{equation}
$\mu$ is an arbitrary constant. $D_{\mu}$ denotes the usual
covariant derivative. To save the place we consider here the group
$SU(2)$.

In the following we shall use the parametrization of $\phi$ in terms
of Hermitean components
\begin{equation}
\phi=(\frac{i \phi^1+ \phi^2}{\sqrt{2}}(1+ \frac{g}{2 \mu} \phi^0),
\frac{\phi^0- i \phi^3(1+g/(2 \mu) \phi^0)}{\sqrt{2}}) \label{3}
\end{equation}
The complex anticommuting scalar fields $b,e$ will be parameterized
as follows
\begin{eqnarray}
b=(\frac{ib^1+b^2}{\sqrt{2}},
\frac{b^0-ib^3}{\sqrt{2}})(1+ \frac{g}{2 \mu} \phi^0)\nonumber\\
e=(\frac{ie^1+e^2}{\sqrt{2}}, \frac{e^3}{\sqrt{2}})\label{4}
\end{eqnarray}
where the components $e^{\alpha}$ are Hermitean, and $b^{\alpha}$
are antihermitean. This particular parametrization of the classical
fields is used as we want to get rid off the ambiguity in choosing
the gauge for quantization completely.

In this parametrization the Lagrangian (\ref{2a}) is invariant with
respect to "shifted" gauge transformations
\begin{eqnarray}
A_{\mu}^a \rightarrow A_{\mu}^a+\partial_{\mu} \eta^a-g \epsilon^{abc}A_{\mu}^b \eta^c\nonumber\\
\phi^a \rightarrow \phi^a + \frac{g^2}{4\mu}\phi^a \phi^b \eta^b+
\mu \eta^a\nonumber\\
\phi^0 \rightarrow \phi^0-\frac{g}{2}\phi^a \eta^a(1+\frac{g}{2}\phi^0)\nonumber\\
b^a \rightarrow b^a+ \frac{g}{2} \epsilon^{abc}b^b \eta^c+
\frac{g}{2}b^0 \eta^a+\frac{g^2}{4\mu}b^a\phi^b \eta^b\nonumber\\
e^a\rightarrow e^a+ \frac{g}{2}\epsilon^{abc}e^b \eta^c+
\frac{g}{2}e^0
\eta^a\nonumber\\
b^0 \rightarrow b^0-\frac{g}{2}b^a \eta^a+\frac{g^2}{4 \mu}(\phi^a \eta^a)\nonumber\\
e^0 \rightarrow e^0-\frac{g}{2}e^a \eta^a.
 \label{5}
\end{eqnarray}
The field $\phi^a$ is shifted by an arbitrary function, therefore
one can put $\phi^a=0$. Contrary to the common wisdom this gauge is
algebraic, but Lorentz invariant. It may be used beyond perturbation
theory as well.

This Lagrangian is also invariant with respect to the supersymmetry
transformations
\begin{eqnarray}
\phi \rightarrow \phi-b \epsilon\nonumber\\
e \rightarrow e- \frac{D^2(\phi- \hat{\mu})}{m^2} \epsilon\nonumber\\
b \rightarrow b \label{6}
\end{eqnarray}
where $\epsilon$ is a constant Hermitean anticommuting parameter.
This symmetry plays a crucial role in the proof of decoupling of
unphysical excitations. It holds for any $\alpha$, but for
$\alpha=0$ these transformations are also nilpotent.

Note that for further discussion we need only the existence of the
conserved charge $Q$
 and nilpotency of the asymptotic charge $Q_0$, as the physical spectrum
 is determined by the asymptotic dynamics.
   In the case under
consideration the nilpotency of the asymptotic charge requires
 $\alpha=0$, and the massive theory with $\alpha \neq 0$ is gauge
 invariant but not unitary. It may seem strange as usually the gauge
 invariance is a sufficient condition of unitarity, because one
 can pass freely from a renormalizable gauge to the unitary one, where
 the spectrum includes only  physical excitations. In the present case
 there is no "unitary" gauge. Even in the gauge  $\phi^a=0$, there are
 unphysical excitations.

For gauge transformations (\ref{5}) the gauge $\phi^a=0$ is
admissible both in perturbation theory and beyond it. Indeed, if
$\phi^a=0$, then under the gauge transformations (\ref{5}) the
variables $\phi^a$ become
\begin{equation}
\delta \phi^a= \mu \eta^a \label{7}
\end{equation}
and the condition $\phi^a=0$ implies that $\eta^a=0$. It is also
obvious that for $\alpha \neq 0$ the Lagrangian (\ref{2a}) describes
a massive vector field and does not produce infrared singularities.

In terms of shifted variables the Lagrangian (\ref{2a}) looks as
follows
\begin{eqnarray}
L=- \frac{1}{4} F_{\mu \nu}^aF_{\mu \nu}^a-m^{-2}(D^2 \phi)^*(D^2
\phi)
+m^{-2}(D^2 \phi)^*(D^2 \hat{\mu})\nonumber\\
+m^{-2}(D^2 \hat{\mu})^*(D^2 \phi)-m^{-2}(D^2 \hat{\mu})^*(D^2 \hat{\mu})+(D_\mu e)^*(D_\mu b)\nonumber\\
+(D_\mu b)^*(D_\mu e)+ \alpha^2(D_\mu \phi)^*(D_\mu \phi)- \alpha^2(D_\mu \phi)^*(D_\mu \hat{\mu})\nonumber\\
- \alpha^2(D_\mu \hat{\mu})^*(D_\mu \phi)+ \alpha^2(D_\mu
\hat{\mu})^*(D_\mu \hat{\mu})- \alpha^2m^2(b^*e+e^*b) \label{10}
\end{eqnarray}

The shift of the variables $ \phi$ produces the term
\begin{equation}
\alpha^2(D_\mu \hat{\mu})^*(D_\mu \hat{\mu})= \frac{\alpha^2
\mu^2}{2}A_ \mu^2 \label{10a}
\end{equation}
which gives a mass to the vector field. The term
\begin{equation}
 m^{-2}(D^2 \hat{\mu})^*(D^2 \hat{\mu})=
\frac{\mu^2}{2m^2}[(\partial_\mu A_\mu)^2+\frac{g^2}{2}(A^2)^2]
\label{11a}
\end{equation}
makes the theory renormalizable for any $\alpha$. To avoid
complications due to the presence of the Yang-Mills dipole ghosts at
$\alpha=0$ we put $\mu^2=m^2$.

Invariance of the Lagrangian (\ref{10}) with respect to the gauge
transformation (\ref{5}) and the supersymmetry transformations
(\ref{6}) makes the effective Lagrangian invariant with respect to
the simultaneous BRST transformations corresponding to (\ref{5}) and
the supersymmetry transformations (\ref{6}). The effective
Lagrangian may be written in the form
\begin{equation}
L_{ef}=L+s_1 \bar{c}^a\phi^a=L(x)+ \lambda^a \phi^a- \bar{c}^a(\mu
c^a-b^a) \label{13}
\end{equation}
One can integrate over $\bar{c},c$ in the path integral determining
expectation value of any operator corresponding to observable. It
leads to the change $c^a=b^a \mu^{-1}$. After such integration the
effective Lagrangian becomes invariant with respect to the
transformations which are the sum of the BRST transformations and
the supersymmetry transformations (\ref{6}) with $c^a=b^a \mu^{-1}$.
These transformations look as follows
\begin{eqnarray}
 \delta A_\mu^a=D_\mu b^a \mu^{-1} \epsilon\nonumber\\
\delta \phi^a=0\nonumber\\
\delta \phi^0=-b^0(1+ \frac{g}{2\mu} \phi^0) \epsilon\nonumber\\
\delta e^a=(\frac{g}{2\mu} \epsilon^{abc}e^bb^c+
\frac{ge^0b^a}{2 \mu}+i\frac{D^2(\tilde{\phi})^a}{\mu^2})\epsilon\nonumber\\
\delta e^0=(-\frac{ge^ab^a}{2 \mu}-\frac{D^2(\tilde{\phi})^0}{\mu^2})\epsilon\nonumber\\
\delta b^a=\frac{g}{2\mu}\epsilon^{abc}b^bb^c\nonumber\\
\delta b^0=0 \label{14}
\end{eqnarray}

For the asymptotic theory these transformations acquire the form
\begin{eqnarray}
\delta A_\mu^a=\partial_\mu b^a \mu^{-1} \epsilon\nonumber\\
\delta \phi^a=0\nonumber\\
\delta \phi^0=-b^0 \epsilon\nonumber\\
\delta e^a=\partial_\mu A_\mu^a \mu^{-1}\nonumber\\
\delta e^0=-\partial^2 \phi^0 \mu^{-2}\nonumber\\
\delta b^a=0\nonumber\\
\delta b^0=0. \label{13a}
\end{eqnarray}

 According to the Neuther theorem the
invariance with respect to the supertransformations mentioned above
generates a conserved charge $Q$, and the physical asymptotic states
may be chosen to satisfy the equation
\begin{equation}
\hat{Q_0}| \psi>_{as}=0 \label{15}
\end{equation}
\begin{eqnarray}
 Q_0= \int d^3x[(\partial_0 A_i^a- \partial_i A_0^a)
\mu^{-1} \partial_i b^a-\mu^{-1}\partial_\nu A_{\nu}^a \partial_0
b^a +\mu^{-2} \partial^2(
\partial_0 \phi^0) b^0- \nonumber\\ \mu^{-2} \partial_0 b^0 \partial^2( \phi^0- \mu
\alpha^2 b^aA_0^a] \label{16}
\end{eqnarray}
Due to the conservation of the Neuther charge this condition is
invariant with respect to dynamics. It was proven in the paper
\cite{Sl4} that this symmetry guarantees the decoupling of all
unphysical excitations at $\alpha=0$ and the transitions between the
states, annihilated by the charge $Q$ include only three
dimensionally transversal components of the Yang-Mills field.
Therefore we succeeded to formulate the Yang-Mills theory in such a
way that in a topologically trivial sector Gribov ambiguity is
absent and the infrared regularization valid beyond perturbation
theory is easily constructed. This approach opens also interesting
possibilities to consider topologically nontrivial sectors and study
the confinement problem.

\section*{Discussion}

A renormalizable manifestly Lorentz invariant formulation of the
non-Abelian gauge theories which allows a canonical quantization
without Gribov ambiguity (including Higgs model) is possible.

In perturbation theory the scattering matrix and the gauge invariant
correlators coincide with the standard ones.

On the basis of this approach infrared regularization of Yang-Mills
theory beyond perturbation theory is constructed \cite{Sl4}

This approach seems to be appropriate for a study of existence of
soliton excitations in Yang-Mills theory. This problem is under
consideration.

\section*{Acknowledgements}
This work was supported in part by the grant for support of Leading
scientific schools NS 46122012.1, grant RFBR 11-01-00296a, and the
Program RAS "Nonlinear dynamics".


\begin{thebibliography}{99}
\bibitem{Sl1} A.A.Slavnov, Phys.Lett.B(1991),391.
\bibitem{Sl2} A.A.Slavnov, Theor.Math.Phys., 161(2009), 204-211.
\bibitem{QS1}A.Quadri,A.A.Slavnov,JHEP, 07(2010),087-109.
\bibitem{QS2}A.Quadri,A.A.Slavnov,
Theor.Math.Phys.,166(2011)291-302.
\bibitem{Sl3} A.A.Slavnov, Theor.Math.Phys.,170(2012),198-202.
\bibitem{Gr}V.N.Gribov,Nucl.Phys.B, 139(1978),1-19.
\bibitem{Si}I.M.Singer, Comm.Math.Phys.,60(1978),7-12.
\bibitem{Zw}D.Zwanziger,Nucl.Phys.B321(1989)591-604;
B323(1989)513-544.
 \bibitem{Sl4}A.A.Slavnov, Theor.Math.Phys.,175(2013),447-453.
\end{thebibliography}
\end{document}